\documentclass[12pt,preprint]{aastex}

\begin{document}

\title{The Evolution of Relativistic Binary Progenitor Systems}
\author{G.J. Francischelli\footnote{francis@mail.astro.sunysb.edu}, 
R.A.M.J. Wijers\footnote{rwijers@sbast3.ess.sunysb.edu}, 
G.E. Brown\footnote{popenoe@nuclear.physics.sunysb.edu}}
\affil{Department of Physics and Astronomy, State University of New York}
\affil{Stony Brook, New York 11794}

\begin{abstract}
Relativistic binary pulsars, such as B1534+12 and B1913+16  are
characterized by having close orbits with a
binary separation of $\sim$ 3 R$_\sun$.  The progenitor of such a
system is a neutron star, helium star binary.  The helium star, with a
strong stellar wind, is able to spin up its compact companion via
accretion. The neutron star's magnetic field is then lowered  to observed
values of about $\sim$ 10$^{10}$ Gauss.  As the pulsar lifetime is 
inversely proportional to its magnetic field, the possibility of observing
such a system is, thus, enhanced by this type of evolution.  We
will show that a nascent (Crab-like) pulsar in such a system can, through
accretion-braking torques (i.e. the ``propeller effect'') and wind-induced
spin-up rates, reach equilibrium periods that are close to observed
values.  Such processes occur within the relatively short helium star 
lifetimes.  Additionally, we find that the final
outcome of such evolutionary scenarios depends strongly on initial
parameters, particularly the initial binary separation and helium star
mass.  It is, indeed, determined that the majority of such systems
end up in the pulsar ``graveyard'', and only a small
fraction are strongly recycled.  This fact might help to reconcile
theoretically expected birth rates with limited observations of 
relativistic binary pulsars.

\end{abstract}

\keywords{binaries: close -- pulsars: general -- stars: neutron}

\section{INTRODUCTION}
\label{Intro}
Two of the four known High Mass Binary Pulsar systems, (HMBP's) --
PSR B1913+16 \citep{hul75} and PSR B1534+12 \citep{wol90}
-- have short orbital periods ($\sim$ 10 hours).
Such systems, upon their eventual mergers, are considered to be important
sources of gravitational wave radiation which may be measured by the next
generation of detectors.  As a result, it becomes desirable to understand
the evolutionary processes that lead to their formation.  
Note that we do not consider PSR B2127+11C \citep{pri91} which
resides in the globular cluster, M15:  globular cluster sources may have
completely different evolution mechanisms which are dominated by dynamical
interactions.  

The link between relativistic binaries and their original O/B main-
sequence progenitor systems are believed to be wide High Mass X-Ray
Binaries (or Be/HMXB's).  The standard evolutionary scenario following the
X-Ray phase \citep{bha91, van93} predicts that the
neutron star enters the hydrogen envelope of its giant companion.  Common
Envelope (CE) evolution ensues as the compact object spirals in, creating
dynamical friction and ultimately expelling the envelope.  The orbit is
then tightened, leaving a helium star, neutron star binary.
However, \citet{che93} showed that a neutron star in CE evolution would likely
form a black hole.  \citet{bro95a} confirmed this scenario, showing that
hypercritical accretion forces $\ge 1 M_{\sun}$ onto the neutron star,
sufficient to form a black hole.  

Brown's alternate scenario for the formation of relativistic binaries
involves, instead, a double helium star binary \citep{bro95a, wet96}.  
If the progenitor O/B
supergiants (ZAMS mass $\sim 24 M_\sun$) are initially very close in mass
(within four percent), the two stars will burn helium at the same time.
Thus it is possible for the neutron star to avoid moving through the
envelope of the secondary.  Although CE evolution takes place, it does so
with two helium stars.  Furthermore, a natural explanation is given as to
why the pulsar, which gains mass by accretion, is the heavier star in the
binary.  Such a result is supported by observations. 
Notice that either scenario leads to a neutron star, helium star binary as
an intermediate stage.  Thus, the evolutionary scenario discussed here is
generally applicable.  

The pulsar born into such a binary is Crab-like with a strong magnetic
field $3-10 \times 10^{12}$ Gauss, and a short spin period, 
30--50 ms.  Small orbital separations (1--3 R$_{\sun}$) and strong helium
star winds ensure heavy accretion onto the neutron star, causing its
magnetic field to lower two orders of magnitude and spinning it up
further.  Competing with wind-fed accretion is the so-called
``propeller effect.''  This mechanism exerts a spin-down torque on rapidly
rotating neutron stars as material is prevented from accreting.  We shall see 
that such a scenario is, indeed, consistent with
observations of relativistic binaries such as PSR 1913+16.  Note that
diminished magnetic field strengths lengthen the observable lifetime 
of a pulsar $(\tau = P^2/2B^2)$.  Thus, an ``observability premium'' is given 
to recycled pulsars.  

In \S2, we discuss our model of the evolution of  a helium
star and neutron star
in a close binary.  Particularly, we examine  how the spin period and
magnetic field of the neutron star as well as the orbital separation of
the system change with time. Since many parameters are involved, an
analytic solution is not readily available.  Instead, in \S3, we discuss
results of a computer code which was set up to analyze the model.
Additionally, since initial conditions are not well understood, we
also discuss the effect of variation of parameters on possible outcomes.
It is determined that the final outcome of the binary evolution strongly
depends on such parameters (particularly initial orbital separation and
helium star mass) and, furthermore, the number of systems which would lead
to an observable, recycled pulsar is highly constrained.

\section{DESCRIPTION OF THE MODEL}
\label{Model}
It has been shown \citep{lan89, woo93} that when a helium
star loses matter from an enhanced stellar wind, the mass-loss rate is
dependent on its total mass. 
Furthermore, it is assumed that when very massive helium stars explode in 
a supernova, there
is a tendency to dissociate the binary.  As we are solely interested in 
modeling progenitors to relativistic binaries, we consider, as a first 
approximation, the lower-bound mass limits from \citet{woo93}, i.e.
$\dot{M}_{\rm He} = -5 \times 10^{-8} M_\sun\, \rm{yr}^{-1}
\left(M_{\rm He}/M_\sun\right)^{2.6}$.  

However, it has recently been argued 
\citep{bro01} that this rate is too high by a factor of 2--3.  This is 
supported by polarization measurements of Thomson scattering in helium stars 
as well as the observed scaling of mass-loss rates with orbital 
period changes \citep{stl93, mof94}.  Using polarization measurements of 
V444 Cygni, $(M_{\rm He} = 9.3 M_\sun)$ \citet{stl93} found $\dot{M}_{\rm He}
= 0.75 \times 10^{-5} M_\sun \, \rm{yr}^{-1}$.  (Note that they use a slightly 
different value for the terminal wind velocity.  See eq.~[\ref{vwind}])  
Extrapolating this rate to all helium star masses and assuming the same scaling
as in Woosley et al., we therefore find:

\begin{equation}
\label{mdothe}
\dot{M}_{\rm{He}}=\left(-2.5 \times 10^{-8}
\frac{M_{\sun}}{\rm{yr}}\right)\left(\frac{M_{\rm{He}}}
{M_{\sun}}\right)^{2.6}
\end{equation}

The helium star nuclear burning lifetime is also dependent on its total
mass.  In an approximation to evolutionary calculations made by
\citet{pac71} and \citet{hab86}, \citet{pol91} estimate 
a functional form for helium star lifetimes as

\begin{equation}
\label{helburnlife}
T_{\rm{He}} = \left\{ \begin{array}{ll}
  (1.148 \times 10^7\:\rm{yr}) \left(\frac{M}{M_\sun}\right)^{-1.6}, &
\mbox{$1.6 < M/M_\sun < 4.8$} \\
(2.37 \times 10^6\:\rm{yr}) \left(\frac{M}{M_\sun}\right)^{-0.6}, & 
\mbox{$M/M_\sun >4.8$}
		\end{array}
	      \right.
\end{equation}
\citet{hab86} has shown that helium stars with  $M \le 2.2 M_\sun$ become
white dwarfs and, thus, are not considered here.  Also note that, as we assume
the progenitor to our model to be a double helium star binary, we only
use half of the total helium star lifetime, given by 
equation~(\ref{helburnlife}),
in our calculations.

We then look to the question of how much of the ejected helium
star matter is actually accreted onto its compact companion ($M_{\rm x}$).
The captured mass rate, $f_{\rm c}$, is defined as the fraction of mass
captured by the neutron star's gravitational field and is given by $f_{\rm
c}$ = $-\dot{M}_{\rm cap}/\dot{M}_{\rm He}$ where $\dot{M}_{\rm cap}$ is
the mass capture rate.  Note that, as $\dot{M}_{\rm He}$ is negative,
$f_{\rm c}$ is always greater than or equal to zero.
Similarly, the accreted mass fraction, $f_{\rm a}$ is given by  the
fraction of mass transferred that is actually accreted onto the surface of
the compact star.  Here, $f_{\rm a}$ = $\dot{M}_{\rm x}/\dot{M}_{\rm
cap}$. As matter can not be accreted faster than the Eddington limit,
we define $f_{\rm a}$ = 1 if $\dot{M}_{\rm cap} < \dot{M}_{\rm Edd}$ but
$f_{\rm a} = \dot{M}_{\rm Edd}/\dot{M}_{\rm cap}$ for super-Eddington
transfers. In sum, we see:

\begin{equation}
\label{transfer}
\dot{M}_{\rm x} = -f_{\rm c}f_{\rm a}\dot{M}_{\rm He}
\end{equation}
Alternatively, one may define the parameter $\alpha$ as the total fraction
of matter that is lost from the system.  In that case, $\alpha = 1 -
f_{\rm c}f_{\rm a}$.

Assuming a standard Keplerian orbit with binary separation, $a$, the
neutron star moves relative to the helium star with a circular orbital
velocity 

\begin{equation}
\label{vorb}
v_{\rm x} = \left(\frac{G(M_{\rm He}+M_{\rm x})}{a}\right)^{1/2}
\end{equation}
In an  approximation to numerical work done by \citet{hab86}
we estimate that the helium star wind velocity at the position of the neutron
star obeys the relation given by,

\begin{equation}
\label{vwind}
v_{\rm w} = v_{\infty}\left(1 - \frac{R_{\rm He}}{a}\right)
\end{equation}
where the helium star radius is given by $R_{\rm He}/R_\sun =
0.22(M_{\rm He}/M_{\sun})^{0.6}$ and $v_{\infty} \approx 2000$ km
s$^{-1}$.
Generally, the helium stellar wind will move out radially,
orthogonal to $v_{\rm x}$, so that the pulsar experiences the wind moving
at a relative velocity of $v_{\rm r} = (v^2_{\rm w} + v^2_{\rm x})^{1/2}.$

Adopting the standard accretion mechanism \citep{bon52}, we
assume incoming matter gets captured near the so-called \emph{accretion
radius}, $R_{\rm g}$.  This is defined as the point where the wind
velocity (relative to the neutron star) is equal to the escape velocity of
infalling matter.  Thus,

\begin{equation}
\label{racc}
R_{\rm g} = \frac{2GM_{\rm x}}{v^2_{\rm r}}
\end{equation}
It should be noted that \citet{bon52} defines the accretion radius as
$R_{\rm g} = 2GM_{\rm x}/(v_{\rm r}^2 + c_{\rm s}^2)$, where
$c_{\rm s}$ represents the speed of
sound in the plasma.  However, for the parameters of this model, this may
be considered negligible and ignored.

Finally, we can use geometry to estimate the captured mass fraction,
$f_{\rm c}$.  We determine it to be the fraction of a sphere of radius,
$a$, which occupies the area enclosed by the accretion radius of the
neutron star, i.e. $f_{\rm c} = \pi R_{\rm g}^2/4 \pi a^2$.  With 
equation~(\ref{racc}), this becomes:

\begin{equation}
\label{fc}
f_{\rm c} = \frac{G^2 M^2_{\rm x}}{a^2\left(v^2_{\rm x} +
v^2_{\rm w}\right)^2}
\end{equation}

\subsection{Orbital Evolution}
\label{orbit}
In order to determine how the orbital separation changes with time, one
needs to model how angular momentum is transferred in the binary.
Neglecting spin angular momentum and assuming circular Keplerian orbits
one finds $J_{\rm orb} = \mu a^2\omega$ where $\omega =\sqrt{G(M_{\rm x}
+ M_{\rm He})/a^3}$ and $\mu$ is the reduced mass.
Recalling that $\dot{M}_{\rm x} = (\alpha - 1)\dot{M}_{\rm He}$ and
differentiating, one can easily show

\begin{equation}
\label{adot}
\frac{\dot{a}}{a} = \frac{2\dot{J_{\rm orb}}}{J_{\rm orb}} -
\frac{2\dot{M}_{\rm He}}{M_{\rm He}}\left[1 + (\alpha -
1)\frac{M_{\rm He}}{M_{\rm x}} -
\frac{\alpha}{2}\left(\frac{M_{\rm He}}{M_{\rm x} +
M_{\rm He}}\right)\right] 
\end{equation}

Since matter is being lost from the system, it is clear that total angular
momentum will not be conserved.  It is then necessary to make some
assumptions about how angular momentum is transferred from one star to
another.  Reasonably, one could expect that the actual mass being
passed from the helium star to the neutron star (i.e. $f_{\rm c}$ 
$dM_{\rm He}$) is transferred conservatively.  We then assume that the
fraction $(1-f_{\rm c}) dM_{\rm He}$  will leave the system with the
specific angular momentum of the
helium star, $\hat{\jmath}_{\rm He}$.  Thus dJ$_{\rm i} =
(1-f_{\rm c})$ $dM_{\rm He} \hat{\jmath}_{\rm He} = (1-f_{\rm c})$ $dM_{\rm He}
(M_{\rm x}/M_{\rm tot})^2a^2\omega$.
Similarly, we expect that $f_{\rm c}(1-f_{\rm a})dM_{\rm He}$ leaves the
system with the specific angular momentum of the neutron star, dJ$_{\rm f}
= f_{\rm c}(1-f_{\rm a})$ $dM_{\rm He} (M_{\rm He}/M_{\rm tot})^2a^2
\omega$. Together, these assumptions yield a differential equation for the
orbital angular momentum of the system, $\dot{J}_{\rm orb} = \dot{J}_{\rm
i} + \dot{J}_{\rm f}$,

\begin{equation}
\label{jorb}
\frac{\dot{J}_{\rm orb}}{J_{\rm orb}} = \frac{\dot{M}_{\rm He}}{M_{\rm x} 
+ M_{\rm He}} \left[(1-f_{\rm c})\frac{M_{\rm x}}{M_{\rm He}} +
f_{\rm c}(1-f_{\rm a})\frac{M_{\rm He}}{M_{\rm x}}\right]
\end{equation}

Defining the mass ratio, $q \equiv M_x/M_{He}$, and combining 
equations~(\ref{adot}) and~(\ref{jorb}), we find

\begin{equation}
\label{adotq}
\frac{\dot{a}}{a} = \frac{\dot{q}\left[q - f_c(2-2q^2 +f_{\rm a} 
q)\right]} {[q+f_{\rm c} f_{\rm a}]q(1+q)}
\end{equation}

\subsection{Emitter Phase}

Initially, the newborn pulsar is characterized by a rapid spin (P= 30 $-$ 50
ms.) and a strong dipole magnetic field ($B_{\rm s} \sim 3-5 \times
10^{12}$ G). Such enhanced dipole radiation pressure might be sufficient
to keep the wind plasma from being accreted onto the neutron star.  In
this phase the compact star will behave like an isolated radio pulsar,
spinning down with time as rotational kinetic energy is converted to
dipole radiation energy.  Such systems have been extensively studied 
\citep{gun69, gol69} and we model dipole radiation pressure for a neutron 
star  of radius $R_{\rm x}$  and spin period, $\Omega$ as 

\begin{equation}
\label{diprad}
P_{\rm rad}(r) = \frac{L}{4\pi r^2 c}  = \frac{B^2R^6_{\rm x}\Omega
^4}{24\pi r^2c^4}
\end{equation}

The \emph{stopping radius}, $R_{\rm s}$, is defined as the point where
dipole radiation pressure is sufficient to balance wind pressure from the
helium star \citep{urp97}.  If the stopping radius is \emph{less} than 
the accretion
radius, then the neutron star will behave like an isolated emitter,
spinning down via the Gunn-Ostriker mechanism.  However, if $R_{\rm g} >
R_{\rm s}$, then it is possible for the pulsar to accrete.  We estimate
wind pressure to be given by $P_{\rm w} \sim
\rho_{\rm w} v_{\rm w}^2$ where $\rho_{\rm w}$, the plasma density, is
approximately given by $\rho_{\rm w} \approx |\dot{M}_{\rm He}|/4\pi a^2
v_{\rm w}$. Finally, setting $P_{\rm w} = P_{\rm rad}(R_{\rm s})$, we get
an expression for the stopping radius

\begin{equation}
\label{rstop}
R_{\rm s} =
\sqrt{\frac{B^2R^6_{\rm x}a^2\Omega^4}{6c^4|\dot{M}_{\rm He}|v_{\rm w}}}
\end{equation}

Note that as no accretion may take place at this point($f_{\rm c} =
f_{\rm a} = 0$), the pulsar magnetic field will remain constant (see 
eq.~[\ref{magdecay}]). The spin period, $P=2\pi/\Omega$ will increase
according to the standard relation \citep{gun69}

\begin{equation}
\label{Gunn}
P\dot{P} = \left(\frac{16\pi^2 R^6_{\rm x}}{3Ic^3}\right)B^2
\end{equation}
Here, I $=k^2 MR^2$ represents the moment of inertia of the neutron
star with $k^2 \sim 0.4$ \citep{pra00}.

\subsection{The Propeller Phase}
\label{propeller}

According to equation~(\ref{rstop}), the stopping radius is proportional to
the inverse square of the period.  Thus, over a relatively short amount of
time, the  pulsar will spin down sufficiently such that accreting matter
may interact with the magnetosphere of the neutron star.  Recall this
occurs at the point where the stopping radius falls inside the accretion
radius.

We assume, for simplicity, that the magnetospheric boundary of the neutron
star is defined where the ram pressure of infalling matter is
balanced by the neutron star's magnetic dipole pressure \citep{lam73}
Assuming spherical inflow (but see Ghosh \& Lamb, 1979a,b), the magnetospheric
radius is, thus:

\begin{equation}
\label{alfven}
R_m = \left(\frac{B^4R_{\rm x}^{12}}{8GM_{\rm x}\dot{M}
_{\rm cap}^2}\right)^{1/7}
\end{equation}

Once matter couples to the neutron star's magnetic field, 
the interaction's effect on the overall spin evolution depends on the
balance between centrifugal and gravitational accelerations.  If the
pulsar rotation is, initially, too fast, the neutron star will eject
infalling plasma, propelling it away tangentially while simultaneously
losing angular momentum in the process. This ``propeller''
effect \citep{pri72, ill75} can be
parameterized by a \emph{fastness parameter}, $\omega_{\rm s} \equiv
\Omega/\Omega_{\rm k}(R_{\rm m})$ where $\Omega_{\rm k}(R_{\rm m}) =
\sqrt{G M_{\rm x}R_{\rm m}}$, is the
Keplerian angular velocity at the magnetospheric boundary.  Clearly, for a
fastness parameter greater than unity, the propeller mechanism is
initiated.  

The resulting angular momentum loss (and associated increase in spin
period) can be represented  by a propeller torque, $N^{\rm J}_{\rm prop} =
I\dot{\Omega} = -\dot{M}_{\rm cap}R_{\rm m} v_{\rm esc}(R_{\rm m})
=-\dot{M}_{\rm cap}\sqrt{2GM_{\rm x} R_{\rm m}}$.  Consequently, the
spin-rate decreases according to the relation:

\begin{equation}
\label{J_spin-down}
\frac{\dot{P}}{P^2} = \frac{\dot{M}_{\rm cap}}{2\pi I} \sqrt{2 G M_{\rm x}
R_{\rm m}}
\end{equation}

It is instructive to note that it is also possible to use energy methods
to determine the propeller torque. Over time, rotational kinetic energy of the
 neutron star will be transmitted through shocks to the wind plasma falling 
near the magnetospheric boundary \citep{fab75}.  Consequently, this gas will 
heat up and be dispersed when it attains escape velocity $(V \sim V_{\rm esc}
 = \sqrt{2 G M_{\rm x}/R_{\rm m}})$.  Thus we find,  $\dot{E} = 
I \Omega \dot{\Omega} = -1/2 \dot{M}_{\rm cap}V_{\rm esc}^2 = 
-GM_{\rm x}\dot{M}_{\rm cap}/R_{\rm m}$.  Hence, $N^{\rm E}_{\rm prop} = 
I\dot{\Omega}=-GM_{\rm x}\dot{M}_{\rm cap}/R_{\rm m}\Omega
= -[GM_{\rm x}\dot{M}_{\rm cap}/R_{\rm m}\Omega_{\rm k}(R_{\rm m})]
\omega_{\rm s}^{-1}.$  

Upon examination of the two possible propeller torques, we see

\begin{equation}
\label{energy_vs_angmom}
N^{\rm E}_{\rm prop} = -\frac{\dot{M}_{\rm cap}}{\omega_{\rm s}}
\sqrt{GM_{\rm x}R_{\rm m}} =
\frac{N^{\rm J}_{\rm prop}}{\sqrt{2} \, \omega_{\rm s}}
\end{equation}
Thus, for extremely fast rotators, it is much more difficult for an energy
propeller to change the spin period of the neutron star significantly.  The
reality of the magnetospheric interaction is very complex, so it is hard to
say which mechanism is inherently more accurate.  Therefore, we use both the 
energy and angular momentum propellers and compare results.

\subsection{Accretion}
\label{accretion}

For fastness parameter less than unity, co-rotating matter is able
to accrete to the surface of the neutron star.  At this point, as 
the helium star wind carries angular momentum, we expect the pulsar spin 
period to decrease over time.  Thus, the pulsar is said to be 
\emph{recycled}.  

It is not initially clear whether spherical or disk 
accretion will dominate over the course of the helium star burning time.  
\citet{sha83} argue that if the intrinsic angular momentum per 
unit mass of accreted gas, $\hat{\jmath}_{\rm acc}$, exceeds the
specific angular momentum of an element in a circular Keplerian orbit 
near the magnetospheric radius, $\hat{\jmath}_{\rm Kep}(R_{\rm m}) =
\sqrt{G M_{\rm x} R_{\rm m}}$, then disk accretion will dominate.  Otherwise,
we may treat the mass-transfer as being (nearly) spherical.  Thus, the 
necessary prerequisite for a disk to form is $\hat{\jmath}_{\rm acc} \ge
\hat{\jmath}_{\rm Kep}$ where it may be shown \citep{sha83},
$\hat{\jmath}_{\rm acc} = (1/2a) v_{\rm x} R_{\rm g}$ and $v_{\rm x}$ and 
$R_{\rm g}$ are given by equations~(\ref{vorb}) and~(\ref{racc}) respectively.

It then becomes desirable to include the effects of magnetic torques
within the disk on the overall spin-rate (viscous
torques may be ignored here).  This complex problem was first discussed
by Ghosh \& Lamb (1979a,b) where they found that for slow rotators
($\omega_{\rm s} \ll 1)$ magnetic coupling may enhance spin-up torques by
as much as forty percent. For $\omega_{\rm s} \la 1$, the
opposite is true and magnetic effects might actually \emph{oppose} the
spin-up.  Following Ghosh \& Lamb,  we define, for disk accretion,
$N_{\rm acc} = n(\omega_{\rm s}) \dot{M}_{\rm x}
\hat{\jmath}(R_{\rm m})$ where the dimensionless coefficient, valid for 
$ 0 \le \omega_{\rm s} \le 0.9$,  is given by

\begin{equation}
\label{ghosh}
n(\omega_{\rm s}) \approx \frac{1.39 \left\{1 - \omega_{\rm s} \left[4.03
\left(1 - \omega_{\rm s} \right)^{0.173} - 0.878 \right]
 \right\}}{\left(1-\omega_{\rm s}\right)}
\end{equation}

Next, we discuss the evolution of the neutron star's magnetic field.
Recent observations and analyses seem to strongly indicate that mass
accretion in binary systems is directly correlated with magnetic field
decay in neutron stars (but see Wijers, 1997) . 
Although there exist many possible accounts of
physical mechanisms that would explain this phenomenon \citep{kon97}, 
here we rely on
the empirical model of \citet{shi89} and leave a more detailed
analysis of accretion driven magnetic field decay for later work.  
For typical initial values of dipole magnetic field strengths, i.e.
$B_{12} \sim 1-10$, it was determined that one can make the following
empirical approximation to magnetic field evolution:

\begin{equation}
\label{magdecay}
B(t) = \frac{B_0}{1 + \frac{\Delta M_{\rm x}}{M_{\rm s}}}
\end{equation}

Here, $M_{\rm s} \sim 12.5 \times 10^{-6} M_\sun$ is a typical scaling
parameter.  This parameter has been set to qualitatively agree with the
observed magnetic fields of Low Mass X-Ray Binaries.  Over their 
lifetimes, LMXB's can accrete up to $\sim 0.1 M_\sun$ and have 
typical magnetic fields on the order of $5 \times 10^8$ Gauss.  Assuming an 
initial field strength of $B_0 \sim 5 \times 10^{12}$ G, we see that 
equation ~(\ref{magdecay}) roughly gives the desired result.

\subsection{Valving and Equilibrium}
\label{valving}

As we have shown, the line of demarcation between spin-up (propeller
effect) and spin-down (accretion) phases of evolution is parameterized by
the fastness parameter, $\omega_s$.  For a very fast rotator, $\omega_s
\gg 1$ and the spin period increases with time as the magnetic field
remains constant.  However, as the neutron star spins down,  matter
co-rotating with the neutron star at the so-called \emph{co-rotation
radius}, $R_{\rm c}$, will spin down as well and $R_{\rm c} \propto
P^{2/3}$. As the
propeller phase continues, the co-rotation radius will increase until it
is finally greater than the magnetopsheric radius, given by 
equation~(\ref{alfven}).  At this point, the neutron star will allow 
co-rotating matter to fall to its surface and the accretion phase begins.  

However, once accreting, the magnetic field begins to decay.  By 
equations~(\ref{alfven}) and~(\ref{magdecay}), we see that 
$R_{\rm m}$ is, roughly,
a monotonically decreasing function of time and 
as accretion continues, the magnetospheric radius diminishes,
possibly allowing the propeller phase to resume.  

In sum, we expect a kind of oscillation or ``valving'' to occur between
accretion and propeller phases of the neutron star's evolution, until
equilibrium is restored.  For spherical mass transfer at equilibrium
the co-rotation radius should coincide with the magnetospheric boundary.
In this case, one can estimate the equilibrium spin period of the neutron star:

\begin{equation}
\label{equilibrium}
P_{\rm eq} = \left(17.3 \ \rm{ms} \right) 
\left(\frac{\dot{M}_{\rm x}}{\dot{M}_{\rm Edd}}\right)^{3/7} B_{10} ^{6/7}
R_6
^{18/7}
  \left(\frac{M_{\rm x}}{M_\sun} \right)^{-5/7}
\end{equation}

So, for a canonical neutron star accreting spherically at the
Eddington limit, $P_{\rm eq} = (13.6 \ \rm{ ms}) B_{10} ^{6/7}$.  As
the magnetic field is brought down with each cycle of accretion, $P_{\rm eq}$
is brought down as well.  

The equilibrium point for disk accretion is generally more complicated.
Upon examination of the Ghosh and Lamb function, (eq.~[\ref{ghosh}]), we
find that $n(\omega_{\rm s})$ is undefined for $\omega_{\rm s} = 1$.  In this 
case, we define the critical ratio, $\omega_{\rm c} \equiv 0.5050$.  
Since $n(\omega_{\rm c}) \sim -1$
and for all $\omega_{\rm s} \ge \omega{\rm c}$ magnetic torques are sufficient
to force the neutron star to spin down, we \emph{define} $\omega_{\rm s} = 
\omega_{\rm c}$ as the turnover from a propeller to an accretor.  I.e., we 
define all $n(\omega_{\rm s} \ge \omega{\rm c}) = -1.$
 
In conclusion, the overall effect of the evolution, over the total helium
burning time, is expected to result in  (a) lowering the neutron star
magnetic field, (b) spinning the pulsar up to or near the millisecond
range, and (c) a slight widening of the orbit.

\section{RESULTS OF THE NUMERICAL CALCULATIONS}
\label{code}

As discussed in the introduction, and expanded on in the previous
section, there are several parameters that need to be accounted for in
order to evolve the binary system with time.  We are, thus,
led to several coupled differential equations for which there is no 
analytic solution.  Instead, we now discuss the results of a computer code
set up to analyze the model.

At each time step within the total helium burning time  
(eq.~[\ref{helburnlife}]), the stopping radius, accretion radius, 
magnetospheric radius and fastness parameter were calculated in order to 
determine the predominant phase of the evolution (i.e. emitter, propeller or 
accretor). The differential equations of \S2 were then solved numerically, 
using a simple Euler scheme, and the calculated parameters were appropriately 
updated. 

Initial conditions  such as stellar masses, neutron star spin period,
magnetic field strength and orbital separation for relativistic binary 
progenitors are largely unknown so the simulation was run for a wide range 
of values.  Specifically, the initial helium star mass was varied in the range
$2.2 M_\sun \le M_{\rm He}\le 15 M_\sun$.  Helium stars with masses below 
the lower limit tend to evolve into white dwarfs \citep{hab86} whereas stars 
with masses far above $15 M_\sun$, will dissociate the binary upon supernova.
Neither possibility would lead to relativistic binary systems. 
It has been shown \citep{bro01} that due to their strong stellar wind, naked 
helium stars in this mass range generally end their lives as neutron stars and
not as black holes.  Similarly, 
as we are interested in the progenitors of close binaries, the initial orbital
separation was chosen to have the (somewhat arbitrary) maximum value of 
$10 R_\sun$.  

The criterion determining the minimum orbital separation was defined such that 
the helium star stay within its Roche lobe, i.e. $R_{\rm L} < R_{\rm He}$.  
If a helium star were allowed to expand beyond its Roche lobe, unstable 
mass-transfer would commence.  Such a scenario would generally not lead to
the formation of a HMBP and is, thus, excluded.  \citet{egg83} has shown that
for a mass-ratio, $q$, (eq.~[\ref{adotq}]) the Roche-lobe radius of a star
is given by 

\begin{equation}
\label{roche}
\frac{R_{\rm L}}{a} = \frac{0.49}{0.6 + q^{2/3} \ln(1 + q^{-1/3})} \equiv
f(q)
\end{equation}

By equation~(\ref{roche}), we see that $a_{\rm min} = R_{\rm He}/f(q)$ and is,
therefore, a function of helium star mass.  Since $R_{\rm He}/R_\sun = 0.22
(M_{\rm He}/M_\sun)^{0.6}$, $a_{\rm min}(2.2 M_\sun) = 0.8435 R_\sun$ and
$a_{\rm min}(15 M_\sun) = 1.914 R_\sun$. 

The initial properties of the nascent neutron star were constrained to a
somewhat smaller range with the Crab pulsar (PSR 0531-21) taken as a
prototype for our model.  Canonical values were given for neutron star
masses, radii and moments of inertia ($1.4 M_\sun, 10^6$  cm,$ \sim
10^{45}$ g-cm$^2$ respectively).  Initial spin periods ranged from 30 -- 50 ms.
and initial dipole field strengths varied on the order $1-10 \times 10^{12}$
Gauss.  

Several plots of the time evolution of spin period and magnetic field under 
various conditions follow and certain general
trends are evident.  A sharp rise in spin period
with time (with no corresponding change in magnetic field) indicates that
the primary factor dominating the evolution is the propeller effect.  
However, when the spin period lengthens to some critical value, the propeller 
mechanism gives way to accretion.  At this point, the period and magnetic 
field strength both decrease with time.

For sufficiently large initial binary separations and/or low helium star
masses, accretion may not occur at all during the total helium burning
time.  In such a case, the dipole field of the neutron star remains
constant and its overall spin period actually \emph{increases} with time,
due to the propeller effect and electromagnetic spindown.  
Many times, this spin-down is sufficiently large to put the neutron star in 
the pulsar graveyard.  

Figure 1 examines the time evolution of the pulsar spin period and 
magnetic field for both the energy and angular momentum propellers 
(see \S 2.3) under typical conditions.  Inital parameters are given by 
$M_{\rm He,i} = 4.0 M_\sun, B_i = 5.0 \times 10^{12} $Gauss, $P_{\rm i} = 
50.0$  ms, and $a_{\rm i} = 1.50 R_\sun$.  The total helium
burning time is calculated, using equation~(\ref{helburnlife}), to be
$6.246 \times 10^5$  years.

Clearly, the final results and overall evolution are sensitively dependent
on the type of propeller mechanism used. For the case of the angular 
momentum propeller, the neutron star quickly enters the accretion 
phase at $t = 2.42 \times 10^4$ years.  Consequently, it ends up with a 
short final period ($99.74$ ms) and low magnetic field strength ($4.169 \times
10^{10}$ G$)$.  Thus, the given initial conditions, coupled with the 
J-propeller mechanism, is sufficient to recycle the pulsar.

Things are quite different for the energy propeller, however.  From Figure
1, we see the neutron star never accretes at all and the 
magnetic field remains constant throughout the evolution.  Weak 
torquing inhibits the rapid spin-down seen for the J-propeller. 
Instead, the period remains roughly constant with $P_{\rm f} =
121.1$ ms.  Additionally, we see that, although the angular momentum propeller
reaches a period close to the equilibrium period (eq.~[\ref{equilibrium}])
early in its evolution $(\sim 4 \times 10^5$ years$)$, the E-propeller never 
does.

For both cases, the helium star loses 0.4848 $M_\sun$ by a steady wind and,
as anticipated, most of this mass is lost.  The neutron star accretes 
$1.494 \times 10^{-3} M_\sun$ for the J-propeller while mass transfer is 
completely non-conservative for the E-propeller.  Note that for this low-mass
helium star, mass-capture is always sub-Eddington and $f_{\rm a} =  1$ for all 
time.  Finally, we see that the type of propeller mechanism has little effect
on the widening of the orbit ($a_{\rm f} = 1.644 R_\sun, 1.645 R_\sun$
respective$)$.  

Figure 2 illustrates the  case where a low-mass helium star ($3
M_\sun$) is in a relatively wide orbit ($a_{\rm i} =  8 R_\sun$).  The 
initial properties of the neutron star are otherwise identical to the previous
case.  Here we see that accretion does not set in until  $t \sim 5.7 \times
10^5$ years for the angular momentum propeller while the energy propeller, 
again, does not accrete at all over the total evolution time of $9.897 \times
10^5$ years.

For the first phase of the evolution, the
weak wind of the helium star is insufficient to push past the stopping
radius of the neutron star.  The pulsar
acts like an emitter, spinning down electromagnetically.  Then at $t \sim
4.9 \times 10^4$ years, the concavity of the curve in the P vs. T diagram
changes. At this point the neutron star has spun down sufficiently for the
helium star wind to begin to interact with its magnetosphere.  However, the
pulsar is still a fast rotator and the propeller phase commences.  

The type of propeller mechanism determines the next stage of the
binary's evolution.  For the case of the angular momentum propeller, the
neutron star rapidly spins down to a maximum pulse period greater than 43
seconds.  At $t = 5.65 \times 10^5$ years, accretion finally begins but
equilibrium is never achieved.  Only $9.880 \times 10^{-6} M_\sun$ is 
actually accreted to the surface $(\Delta M_{\rm He} = -0.3636 M_\sun)$
and the final spin period and magnetic field
strength of the neutron star are 33.14 seconds and $2.798 \times 10^{12}$
Gauss respective.  Clearly the pulsar has entered the graveyard and this
system would not be observed as a relativistic binary.  

Once again, for the energy propeller, weak torquing prevents significant
changes
in the spin period and, although the neutron star never accretes at all
(magnetic field remains unchanged), the final spin period is 298.9 ms.  As
a result, the neutron star can be observed as a radio pulsar (yet it is
clearly not recycled).  For both scenarios, the final orbital separation
is $8.720 R_\sun$. 

Finally, we examine the case where a massive helium star lies in a close
orbit with a neutron star. In this case we expect there to be heavy
accretion and recycling.  In particular, consider a $12 M_\sun$ helium
star in orbit with a canonical neutron star ($P_{\rm i} = 50$  ms, 
$B_{\rm i} = 5 \times 10^{12}$ G) with an initial binary separation of 
$3.0 R_\sun$.

Figure 3 illustrates the resulting evolution of spin and magnetic
fields for the angular momentum propeller.  We see there is, indeed, heavy
accretion and after a sharp propeller cycle, equilibrium is achieved at
$P_{\rm f} = 60.41$ ms.  Additionally, the neutron star accretes $2.086 \times
10^{-3} M_\sun$ which is sufficient to bring the field down to $2.993
\times 10^{10}$ Gauss in a time of $2.668 \times 10^5$ years.  The final
orbital separation is $3.833 R_\sun$.  

Mass transfer is spherical throughout the evolution and,
due to the high initial mass of the helium star, wind loss is super-Eddington 
until $t = 2.54 \times 10^5$ years.  The minimum period of 59.35 ms occurs
near $t \sim 2.1 \times 10^5$ years where steady accretion gives way to
valving.  At this point, the period slowly increases and closely tracks the 
(time-dependent) equilibrium period.

The energy propeller again prohibits accretion
onto the neutron star.  Through electromagnetic spindown and 
propeller torque, the pulsar gradually increases its spin to a final period of 
83.71 ms.  All of the helium star wind matter ($2.944 M_\sun$) is lost
while the magnetic field remains constant.  

For sufficiently heavy accretion, it is possible that slight changes in 
initial 
orbital separation may result in significant differences in the neutron star's
spin period evolution.  Consider the case of a $10 M_\sun$ helium star in orbit
about a neutron star with initial magnetic field $7.5 \times 10^{12}$ G, and 
initial spin period $35$ ms.  
Assume an angular momentum propeller effect only.
Figure 4 shows the resulting time evolution for
$a_{\rm i} = 1.95 R_\sun$ and $2.05 R_\sun$ respectively.  We find that
the binary in the closer initial orbit actually has a \emph{longer} final spin
period.  Such a result seems counterintuitive and needs to be examined more
closely.

When $a_{\rm i} = 2.05 R_\sun$, accretion is spherical 
throughout the mass transfer phase, lasting from $t = 6.1 \times 10^3$ years 
until the time of the helium star's supernova at $t=2.98 \times 10^5$ years.  
The final, equilibrium spin period is 48.4 ms.  However, for the case when 
$a_{\rm i} = 1.95 R_\sun$, accretion changes from disk-type to spherical.
The short, initial, electromagnetic spin-down phase is nearly identical for 
both cases, as is the initial propeller phase.  Then, at $t \sim 5.6 \times
10^3$ yr, spherical accretion begins.  This continues, uninterrupted until 
$t \sim 4.2 \times 10^4$ years.  At this point, a second, smaller, propeller 
phase begins and continues until $t = 4.25 \times 10^4$ years.  Valving and 
disk accretion sets in at this time, allowing the spin period to 
gradually decrease with time.  Finally, at $t = 2.46 \times 10^5$ years, 
$\hat{\jmath}_{\rm acc}$ falls below $\hat{\jmath}_{\rm Kep}$ and spherical 
accretion
commences again.  This results in a much faster drop in spin-period with time.
The neutron star spherically accretes until the helium star supernova and the
final spin period for this case is 55.0 ms.

\section{DISCUSSION}
\label{discussion}

Next we examine the results of a contour plot which parameterizes the
final spin period and magnetic field as functions of initial orbital
separation and helium star mass.  Standard initial values were again
chosen for the neutron star with $M_{\rm x,i} = 1.4 M_\sun$, 
$B_{\rm i} = 5 \times 10^{12}$ G, and $P_{\rm i}=50$ms.  As discussed in
\S3, the initial helium star mass range is $2.2 M_{\sun} \le M_{\rm He,i}
\le 15 M_\sun$ whereas the initial orbital separation varied in the range
$a_{\rm min} \le a_{\rm i} \le 10 R_\sun$.  Here, $a_{\rm min}$ is defined to
be $R_{\rm He}/f(q)$ (eq.~[\ref{roche}]).  The output of the contour plots 
representing the angular momentum propeller and energy propeller 
appear in figures five and six, respectively.  

From the slope of the contours in Figure 5a, it is clear that, for close 
orbits, the final spin period most strongly depends on the initial orbital 
separation, and less on the helium star mass.  However, for wider (initial) 
orbits, both parameters play an important role on the final outcome. 

If neutron stars act as angular momentum propellers, then
for any orbit with initial separation greater than $\sim 6 R_\sun$, the pulsar
will not only be unrecycled but it will sit in the graveyard as
well.  Therefore, there is a strong constraint placed on progenitors of
relativistic binary pulsars such as 1913+16.

Recall that pulsar spin-down times are proportional to $P^2/B^2$.
Thus, one expects an ``observability premium'' for low-field, fast
rotators. I.e., due to longer lifetimes outside the graveyard, one should
expect to observe a higher proportion of such systems.  Indeed this is
true of the two relativistic binaries (three including 2127+11C).  It
seems that, upon examination of figure 5a, that
in order to bring the field below $5 \times 10^{10}$ Gauss $(0.01 B_{\rm i})$, 
the initial binary separation must be a \emph{maximum} of $\sim 4  R_\sun$.
Additionally, if one wishes to observe a system with spin periods on the
order of 50 ms, even closer orbits are necessary, with $a_0 \sim 2 - 3
R_\sun$.  And this is only true for the most massive of helium stars,
i.e. those with $M_{\rm He,i} > 8 M_\sun$.  Thus for angular momentum type
propellers, we conclude that only a minute portion of the overall
parameter space represented in figure four will produce relativistic
binary pulsars such as PSR's 1913+16 and 1534+12. 

The parameter space of final outcomes is much more highly constrained
for the energy propeller (figures 6a,6b) than for the angular momentum
propeller.  It seems, from examining the spin period contour plot in
figure 6a that it is extremely difficult (if not impossible) for a neutron
star to spin down into the graveyard by the E-propeller mechanism.  Even
in the weak-wind, wide orbit limit, $P_{\rm f}$ never rises much higher
than $\sim 0.5$ seconds.  At the other extreme, we find that it is
also very difficult to force the neutron star to accrete at all for this
scenario.  Figure 6b shows that mass transfer, parameterized by magnetic
field decay, occupies only a tiny portion of M-a space.  

From equation~(\ref{energy_vs_angmom}), we see that the propeller torque is
inversely proportional to the fastness parameter.  Thus, unlike the
J-propeller, fast rotators are unable to easily change their spin period.
Thus, one may conclude that the energy propeller is highly inefficient
at recycling pulsars.

This characteristic becomes a significant factor in determining whether an
E-propeller will accrete at all or not.  As with the J-propeller, slightly
changing the initial conditions (particularly the initial orbital
separation) may make a profound difference in the final outcome of the
pulsar's evolution (although, for different reasons).  The result of this
condition is the dense band of contours near the $a_{\rm min}$ line in
figure 6a.

For a specific example of why this occurs, consider a $6
M_\sun$ helium star orbiting a neutron star with initial parameters
$B_{\rm i} = 4 \times 10^{12}$ G and $P_{\rm i} = 70$ ms.  Figure 7a
shows a plot of final spin period as a function of initial separation in the 
range $a_{\rm min} = 1.27 R_\sun \le a_{\rm i} \le 2.5 R_\sun$.  
A sharp peak exists at the point, $a_{\rm i} = 1.567 R_\sun$.  
Here, $P_{\rm f}$ reaches a maximum value of  3.063
seconds.  For $a_{\rm i} > 1.6 R_\sun$,
$P_{\rm f}$ falls dramatically to a level of about $\sim 85$ ms at
$a_{\rm i} = 2.1 R_\sun$.  At this point there is a much more gradual
increase in final spin period with increasing separation.  

What causes the sudden shift in final spin period?  As figure 7b shows, a
slight change in initial orbital separation will have an effect on the 
fastness parameter and, thus, the overall evolution of the propeller phase.
Here, we examine the time evolution of the spin period for the three
different initial orbital separations -- $1.525 R_\sun, 1.565 R_\sun$,
and $1.605 R_\sun$.  As the initial orbital separation increases, the 
propeller torque will decrease by a small amount and, as a result, it takes 
longer for the neutron star to reach its maximum period.  For $a_{\rm i} =
1.525 R_\sun$, the spin period reaches a maximum of 3.09 seconds at $t =
3.43 \times 10^5$ years.  Afterwards, the accretion phase begins and
continues until the end of the helium star lifetime at $4.044 \times 10^5$
y. where $P_{\rm f} = 177.7$ ms.  
A relatively long accretion phase allows the magnetic field to decay
to a final value of $1.358 \times 10^{11}$ G.  So, we see a slightly
recycled neutron star for the given initial conditions.

Changing the initial separation to $1.565 R_\sun$ increases the fastness
parameter and, consequently, decreases the propeller torque.  Now the
peak does not occur until nearly the end of the evolution $(t =
4.03 \times 10^5$ years$)$.  As there is little time for the accretion
phase, the magnetic field does not change much $(B_{\rm f} = 2.6 \times
10^{12}$ G$)$ and the final spin period is 2.33 seconds. Thus, the pulsar
winds up in the graveyard and is not observable.   Finally, by
increasing the initial separation to $1.605 R_\sun$  we can completely
eliminate the accretion phase altogether.  Here, the pulsar peak never
occurs and the neutron star can not be recycled.  The magnetic field
remains constant throughout the evolution $(P_{\rm f} = 234.9$ ms$)$.

Next we consider the case of PSR J1518+4904 (Nice et al., 1996).  Of all four 
known HMBP's, this one is the most recycled $(P=40.94$  ms, $ \log B = 9.1)$.  
However, it resides in a relatively wide binary with an orbital period of
$8.634$ days $(a \sim 25 R_\sun)$.  Unfortunately, our model can not account
for such properties.

From figure 5a (J-propeller), we see that if $a_{\rm i} > 6 R_\sun$, the 
pulsar is destined to end up in the graveyard.  Even for the most massive 
helium star companions, the neutron star will never accrete and, instead, a 
large propeller torque will sharply increase its spin period.  Such a trend
diminishes for $a_{\rm i} \gg 10 R_\sun$ since, for very wide binaries, 
$R_{\rm g} > R_{\rm s}$ for most (if not all) of the evolution.  The helium
star wind is too weak to interact with the neutron star's magnetosphere at 
such separations, and the pulsar spins down by the Gunn-Ostriker mechanism.
Thus, we would expect that J1518+4904 would not be observable nor would it 
undergo magnetic field decay.

Although the evolutionary history of PSR 1518+4904 is unknown, it is clear that
our model does not account for all possibilities.  We, therefore, must assume 
its evolution was different than the other HMBP's.  One possible evolutionary 
scenario which may account for the observed properties of 1518+4904 is reverse
case C mass-transfer \citep{kip67}.  As low-mass helium stars tend to have 
significantly extended envelopes \citep{hab86}, even an initially wide binary
may undergo sufficient mass-transfer to initiate recycling during the helium 
shell-burning stage.  Assuming the neutron star survived the resulting 
spiral-in phase, it may be possible to end up with a recycled pulsar in 
a wide binary.  Thus, we find that two possible improvements to our current 
model would be an accurate treatment of case C mass-transfer and a more 
detailed physical model of magnetic field decay.  

\section{CONCLUSION}
\label{conclusion}

Relativistic binary pulsars, such as B1534+12 and B1913+16 are characterized by
having close orbits $(a \sim 3 R_\sun)$, with recycled pulse periods and 
magnetic fields $(P \sim 30 - 60$ ms, $\log B \sim 10)$.  A single case of a 
wide HMBP (B1518+4904) exists.  We do not consider PSR 2127+11C as it resides 
in the globular cluster, M15 and may have a different evolutionary history.  

We assume wind-fed mass transfer is responsible for recycling the observable 
neutron star in a HMBP.  In lieu of a more detailed physical mechanism, we use 
the empirical model of \citet{shi89} to account for magnetic field 
decay.  Such a model assumes changes in the pulsar magnetic field are 
proportional to the amount of of mass accreted to its surface.
	
We find that for close initial orbits $(a_{\rm i} \sim 2 - 3 \,R_\sun)$ and an 
initial helium star mass in the range $8 M_\sun \la M_{\rm He,i} \la 
15 M_\sun$, we are able to reproduce the observed spin periods, orbital 
separations, and magnetic fields for HMBP's 1913+16 and 1534+12.  \citet{bro01}
find that helium stars in this mass range will generally end their lives as 
neutron
stars.  Whereas such high mass He stars are relatively rare, they do 
not undergo a significant red giant phase during helium shell burning.  
Such a phase would, in fact, make it more difficult to form a binary pulsar 
for the less massive systems because reverse case C mass-transfer
would generally lead to the formation of a black hole \citep{fry97}.
Finally, we note that with such a high initial helium star mass,
a kick velocity upon supernova would be required if the system is to remain 
bound.  This is consistent with earlier calculations made by \citet{bur86}.

Our model was unable to account for the properties of PSR 1518+4904  
and we speculate that another mechanism such as reverse case C mass-transfer 
is responsible for its properties.  

Cycles of accretion coupled with the propeller effect allow the neutron star
to come to be recycled in a time consistent with helium star nuclear lifetimes.
Of the two possible propeller mechanisms proposed, only the angular momentum 
propeller is efficient at recycling the HMBP progenitors to observed 
properties.

As pulsar lifetimes are inversely proportional to their magnetic fields
$(\tau \sim P^2/B^2)$, recycling and magnetic field-decay lengthen the 
observable lifetime for a neutron star.  Thus, an ``observability premium'' is 
introduced for relativistic binary pulsars.
It is, furthermore, discovered that the  final outcome of the binary evolution
strongly depends on initial conditions (particularly $M_{\rm He,i}$ and
$a_{\rm i})$.  Thus, the number of possible observable systems are
constrained to a small region of the overall parameter space of initial
conditions.

\acknowledgements

We wish to thank Thomas Tauris for many useful discussions.  This work is
partially supported by the U.S. Department of Energy under grant no. 
DE-FG02-88ER40388.

\clearpage

\begin{figure}
\figurenum{1}
\epsscale{0.4}
\plotone{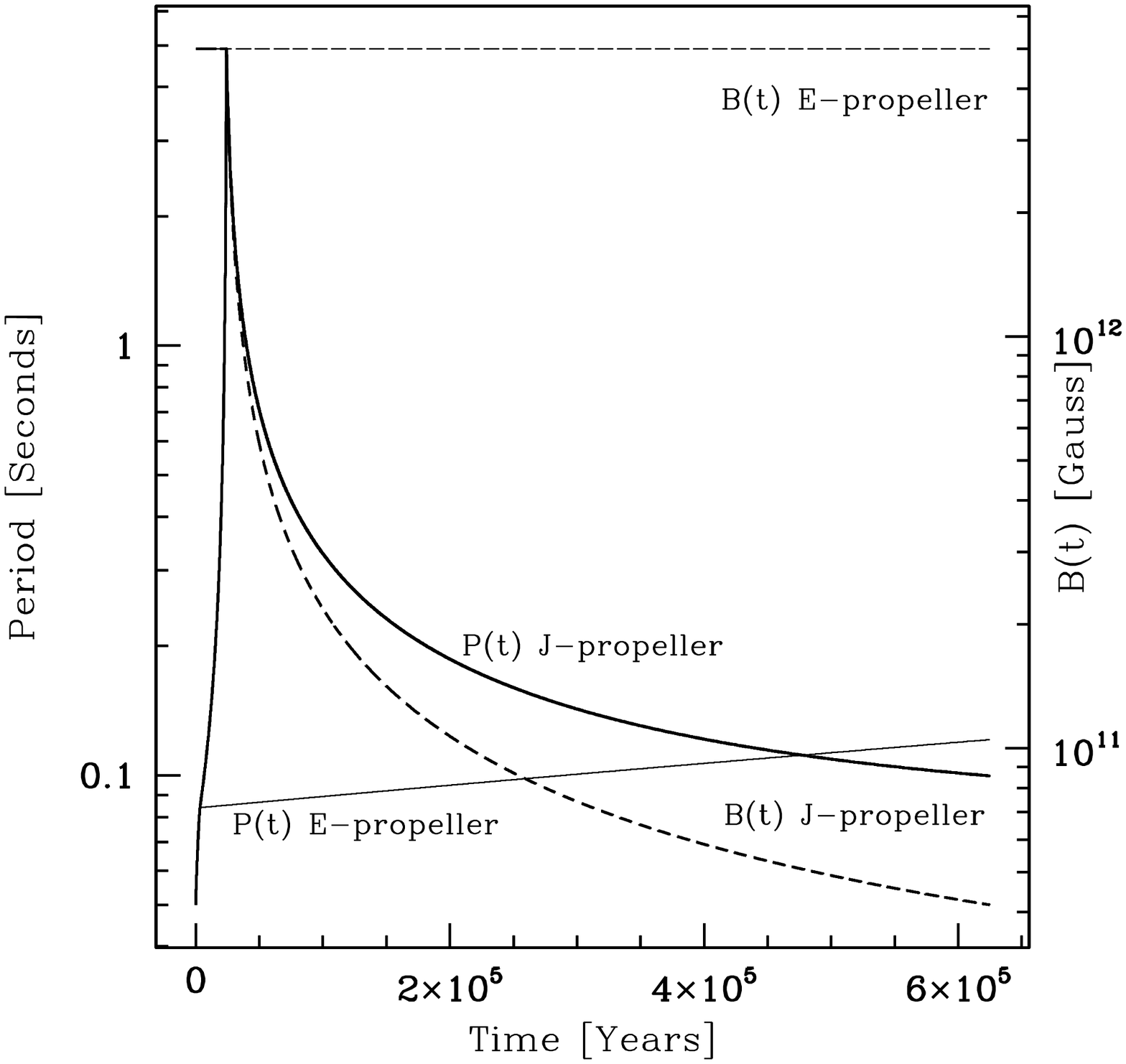}
\caption[]
{\footnotesize Time evolution of the pulsar spin period (solid lines) and 
magnetic field (dashed lines) for both the angular momentum and energy 
propellers. For both cases, 
$B_{\rm i}=5 \times 10^{12}$ G, $P_{\rm i} = 50$ ms, and  $a_{\rm i}=1.50 R
_\sun$.  Initial helium star mass is $4.0 M_\sun$. For the J-propeller, we see
evidence of recycling with $P_{\rm f} = 99.74$ ms and $B_{\rm f} = 4.169
\times 10^{10}$ Gauss.  The neutron star accretes $1.494 \times 10^{-3} M_\sun$
over an evolution time of $6.246 \times 10^5$ years.  For the E-propeller, weak
torquing inhibits a rapid spin-down and, consequently, there is no accretion.  
The final spin period is $121.1$ ms while the magnetic field does not change.
}
\end{figure}


\begin{figure}
\figurenum{2}
\epsscale{0.4}
\plotone{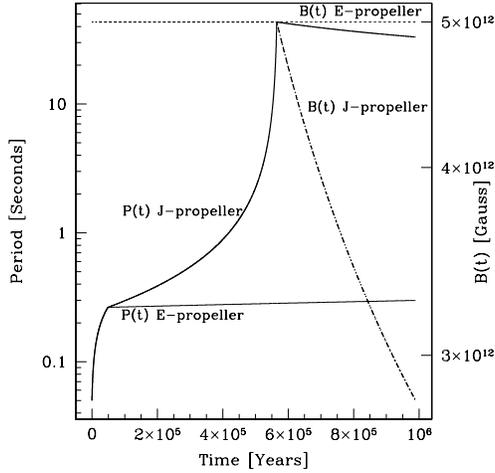}
\caption[]
{\footnotesize Time evolution for the case of a low-mass helium star $(3 M_\sun
)$ in a wide orbit $(a_{\rm i}= 8 R_\sun)$.  Other initial conditions are the 
same as in figure 1.  For the J-propeller, the pulsar will end up in the 
graveyard with a final spin period of 33.1 seconds.  Only $9.88 \times 10^{-6}
M_\sun$ is accreted onto the surface of the neutron star $(\Delta M_{\rm He} =
-0.3636 M_\sun)$ and there is little field decay $(B_{\rm f} = 2.798 \times 
10^{12}$ Gauss$)$.  For the E-propeller, once again, there is no accretion as
evidenced by the constant magnetic field.  Here, the final spin period is 
298.9 ms and the neutron star is observable as a radio pulsar.  For both 
scenarios, the total evolution time is $9.897 \times 10^5$ years and the orbit
widens until $a_{\rm f} = 8.720 R_\sun$.
}
\end{figure}

\clearpage

\begin{figure}
\figurenum{3}
\epsscale{.4}
\plotone{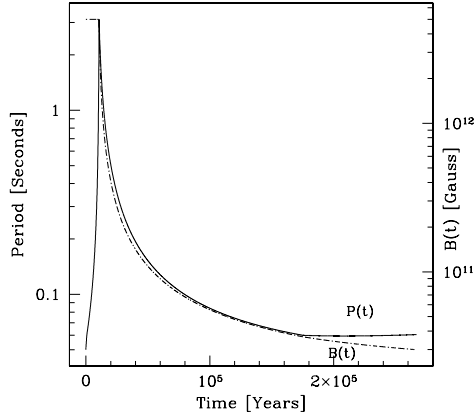}
\caption[]
{\footnotesize Spin and magnetic field evolution for the case $a_{\rm i} =
3 R_\sun, M_{\rm He,i} = 12 M_\sun$ (angular momentum propeller only).  All 
other initial conditions are the same as in previous cases.  There is heavy 
accretion and, after a sharp propeller cycle, the neutron star spins up to an
equilibrium period of 60.41 ms.  The field strength is, subsequently, lowered 
to a final value of $2.993 \times 10^{10}$ Gauss in a time of $2.668 \times
10^5$ years.  The final orbital separation is $3.833 R_\sun$.  For the 
E-propeller (not shown), there is no recycling and $P_{\rm f} = 83.71$ ms.
}
\end{figure} 

\begin{figure}
\figurenum{4}
\epsscale{0.4}
\plotone{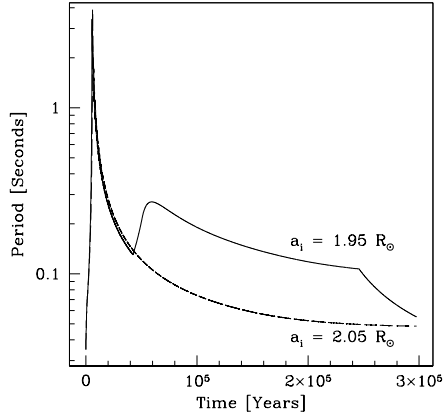}
\caption[]
{\footnotesize Here we see the time evolution of the spin period for a 
canonical neutron star $(P_{\rm i} = 35$ ms, $B_{\rm i} = 7.5 \times 10^{12}$ G
$)$ in a close orbit with a $10 M_\sun$ helium star (J-propeller).  For 
$a_{\rm i} = 2.05 R_\sun$. there is continuous spherical accretion from 
$t=6.1 \times 10^3$ years until $t_{\rm f}= 2.98 \times 10^5$ years.  The 
final equilibrium spin period is 48.4 ms.  However, by slightly changing the 
initial conditions, such that $a_{\rm i} = 1.95 R_\sun$, the time evolution
becomes quite different.  Here, spherical accretion dominates from $5.6 \times 
10^3$ yr. until $4.2 \times 10^4$ yr.  This leads to a second (smaller) 
propeller phase and a cycle of disk accretion.  Finally, at $t=2.46 
\times 10^5$ years, accretion again becomes spherical and $P_{\rm f} = 55$ ms.
}
\end{figure}

\clearpage

\begin{figure}
\figurenum{5}
\epsscale{1.3}
\plottwo{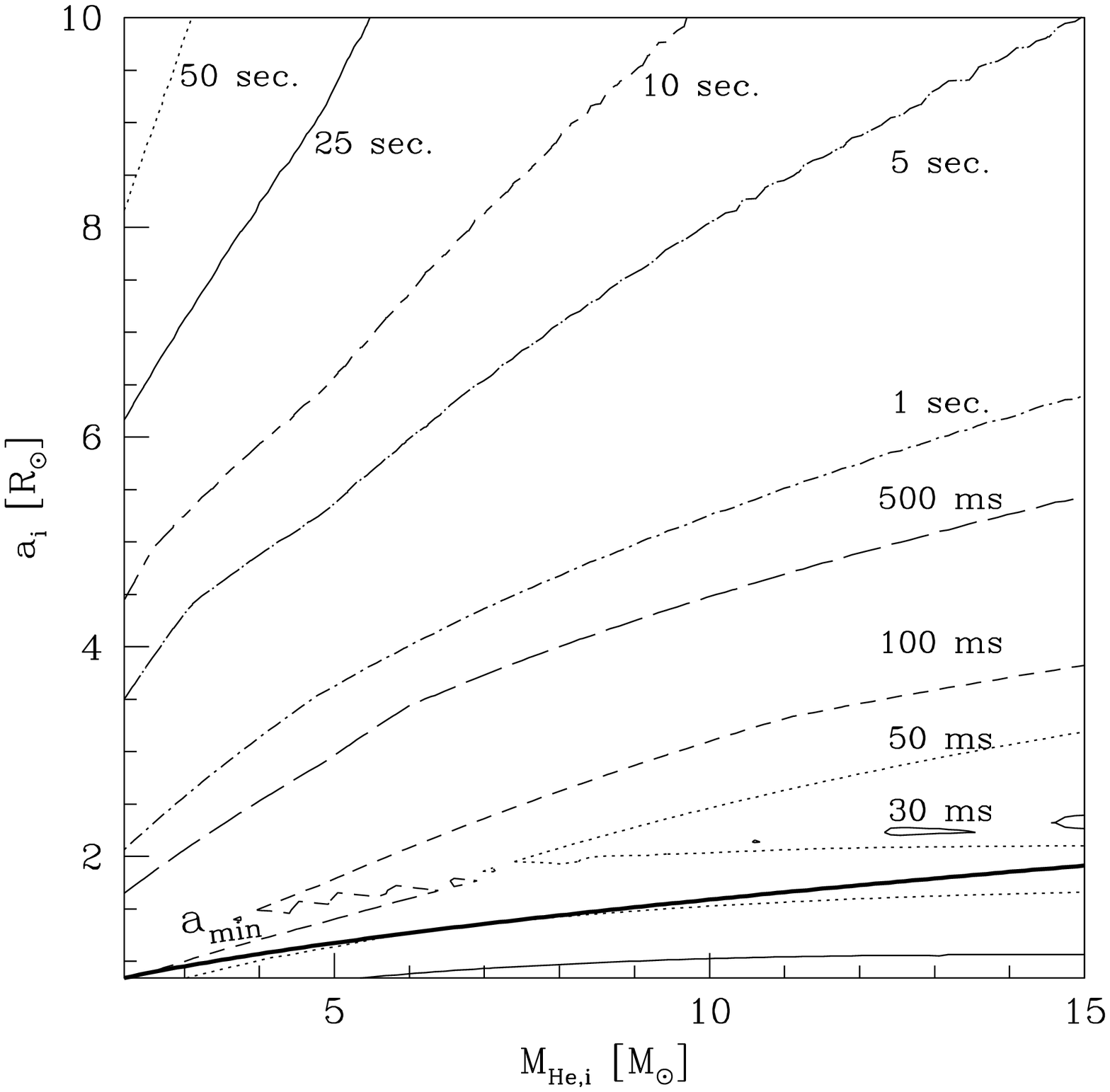}{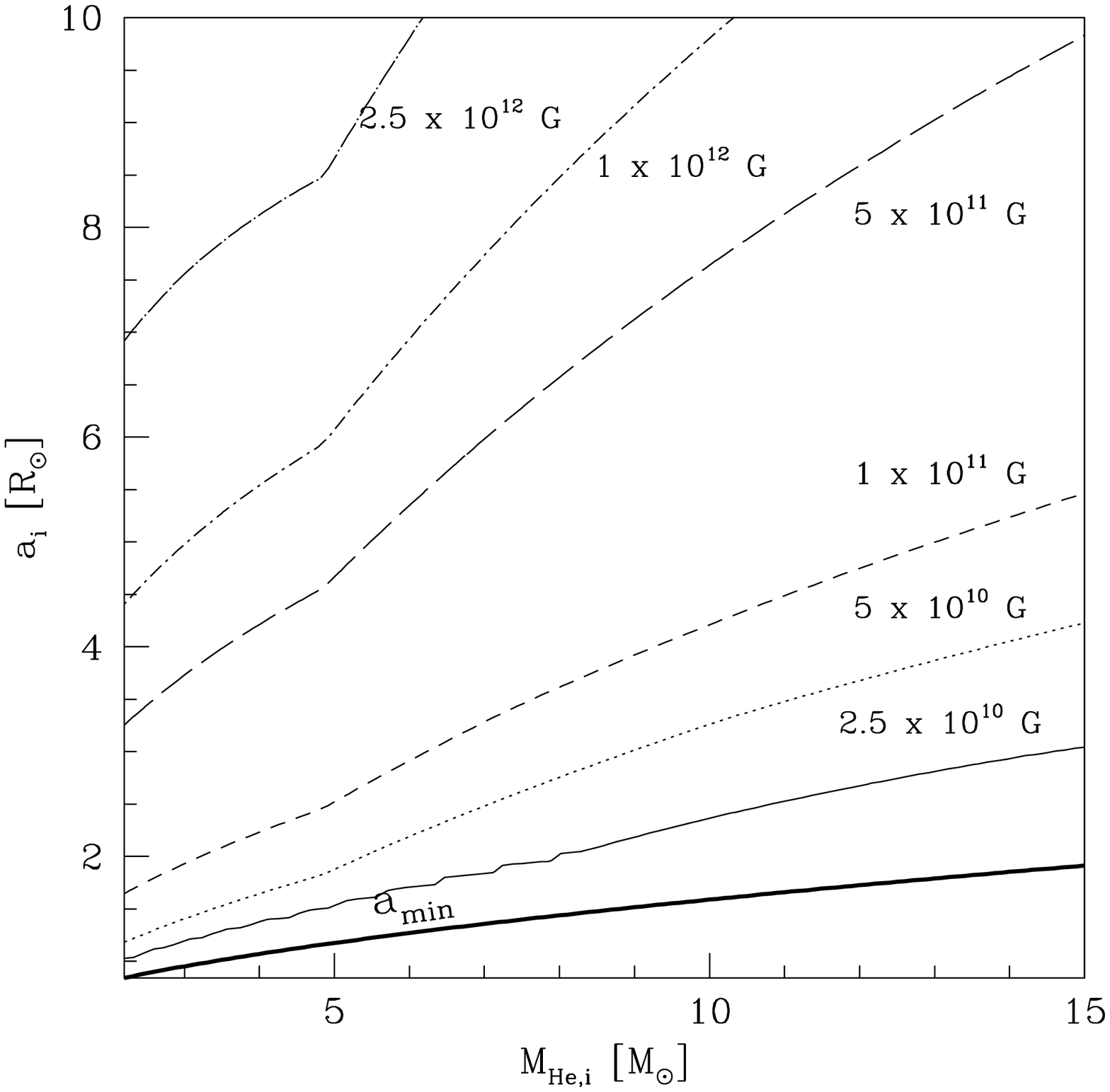}
\caption[]
{\footnotesize Contour plots of (a) final spin period and (b) final magnetic 
field strengths as a function of initial separation and helium star mass for 
the angular momentum propeller effect.
Initial values are given using the Crab pulsar (PSR 0531-21) as a prototype.
Therefore, we have set $B_{\rm i} = 5 \times 10^{12}$ Gauss and $P_{\rm i} = 
50$ ms.  Variation of these parameters do not change the final outcome much.  
The initial helium star mass varied in the range $2.2 \le M_{\rm He} \le
15 M_\sun$ and initial orbital separation range was $a_{\rm min} \le a \le
10 R_\sun$. The $a_{\rm min}$ line is defined by $a_{\rm min} \equiv
 R_{\rm He}/f(q)$ (\emph{see text}).   The lower right portion of the graph 
indicates heavy recycling whereas the upper left contours represent neutron 
stars that have spun into the graveyard.
}
\end{figure}

\clearpage


\clearpage

\begin{figure}
\figurenum{6}
\epsscale{1.3}
\plottwo{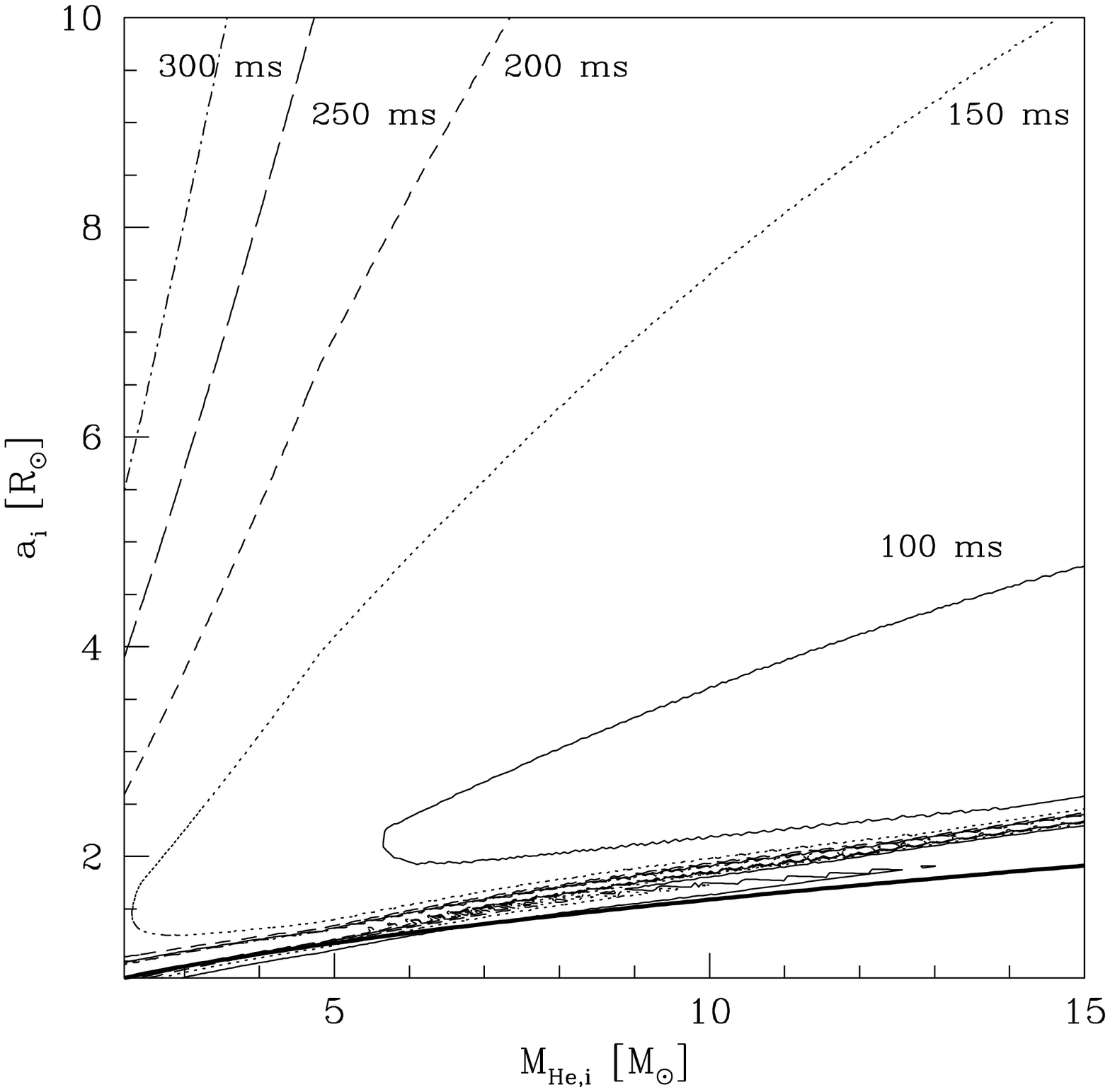}{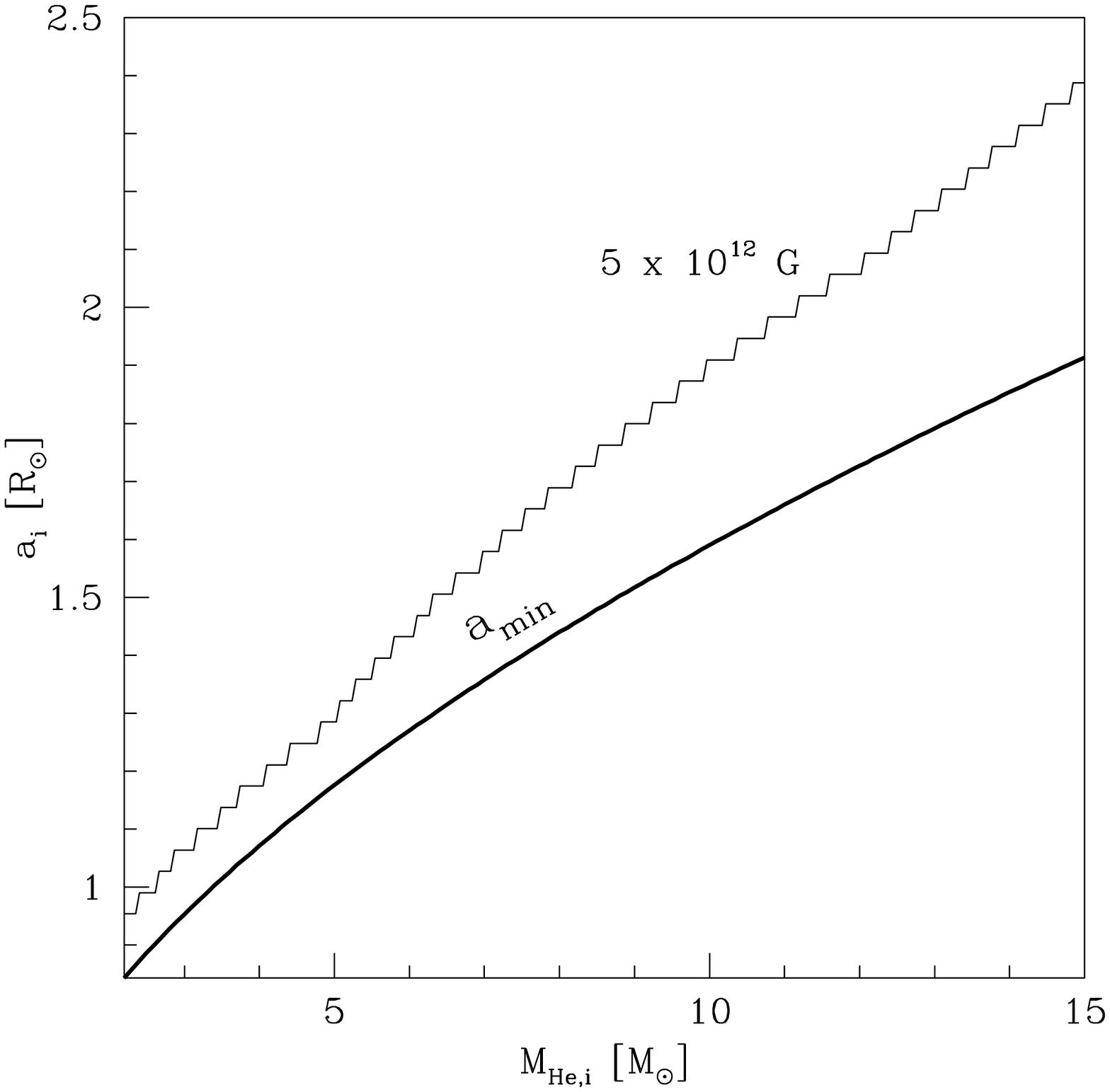}
\caption[]
{\footnotesize Contour plots of (a) final spin period and (b) final magnetic
field as functions of initial orbital separation and helium star mass for the
energy propeller.  Initial conditions and parameter space are the same as for
the angular momentum propeller.  Note that final values are much more strongly
constrained and, it seems, it is very difficult for the E-propeller mechanism 
to either spin the neutron star into the graveyard or sufficiently recycle it 
to form a relativistic binary such as PSR 1913+16.  In figure (6a) we see that
final spin periods near the $a_{\rm min}$ line are sensitively dependent on the
initial conditions (\emph{see text for an explanation}).  From (6b) we find 
that accretion, and, subsequent magnetic field-decay is absent for the vast 
majority of the parameter space in the case of the E-propeller.  
}
\end{figure}

\clearpage

\begin{figure}
\figurenum{7}
\epsscale{1.3}
\plottwo{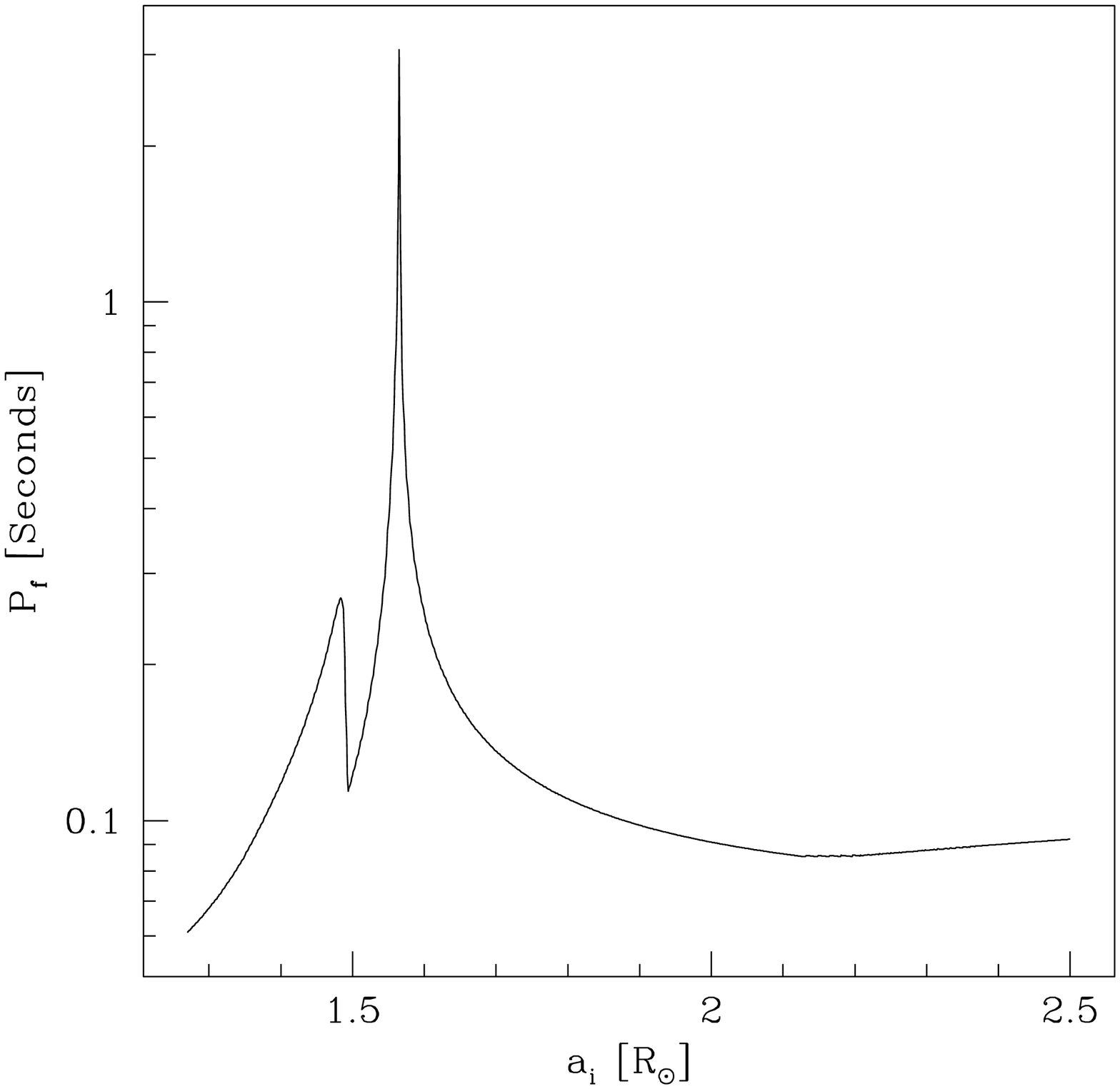}{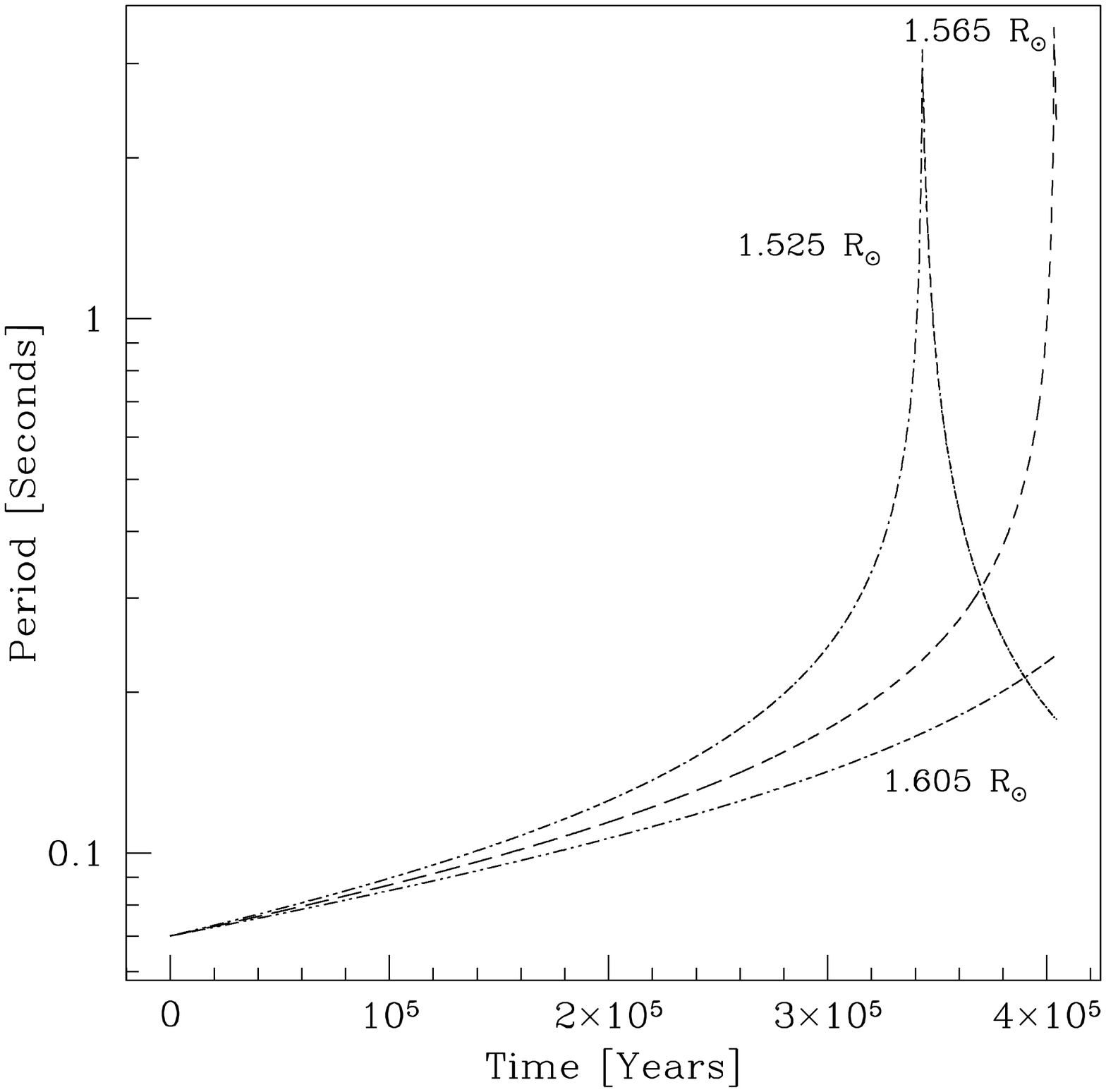}
\caption[]
{\footnotesize (a) A plot of final spin period as a function of initial 
separation
in the range $1.27 \le a_{\rm i} \le 2.5 R_\sun$ (Energy propeller, 
$B_{\rm i} = 4 \times 10^{12}$ G, $P_{\rm i} =70$ ms, $M_{\rm He,i} = 6 M_\sun$).  A sharp 
maximum occurs at the point $a_{\rm i} = 1.567 R_\sun$ where $P_{\rm max} = 
3.06$ seconds.  For $a_{\rm i} > 1.6 R_\sun$, $P_{\rm f}$ falls dramatically
to about $\sim 85$ ms at $a_{\rm i}= 2.1 R_\sun$.  At this point, the final
period gradually rises again with increasing separation.  (b) A plot of the 
time evolution of the spin period for three different initial orbital 
separations -- $1.525 R_\sun$, $1.565 R_\sun$, and $1.605 R_\sun$.  See text
for details.   
}
\end{figure}

\end{document}